\documentclass[twocolumn,amsmath,amssymb,aps,prb,superscriptaddress]{revtex4-2}
\usepackage{epsf}
\usepackage{graphicx}
\usepackage{sidecap}
\usepackage{soul}
\usepackage{array}
\usepackage{amsmath}
\usepackage{amssymb}
\usepackage{dcolumn}
\usepackage{epstopdf}
\usepackage{bm}
\usepackage{amsmath}
\usepackage{physics}

\usepackage{gensymb}
\usepackage{color}
\usepackage{hyperref}
\usepackage{soul}
\sethlcolor{green}
\usepackage{bbding}
\usepackage{setspace}
\hypersetup{
    colorlinks=true,
    citecolor=red,
    linkcolor=red,
    filecolor=blue,   
    urlcolor=blue,
}

\usepackage[normalem]{ulem}

\usepackage{graphicx,color}
\usepackage{amsfonts}
\usepackage[figuresright]{rotating}  
\usepackage{amssymb}
\usepackage{amsmath}

\usepackage{bbold}
\usepackage{wrapfig,lipsum,booktabs}
\usepackage{mathtools}
\usepackage{psfrag}
\usepackage{subfigure}
\usepackage{multirow}
\usepackage{tabularx}
\usepackage{textcomp}
\usepackage{bm}
\usepackage{hyperref}
\usepackage{relsize}
\hypersetup{
 pdfnewwindow=true, colorlinks=true,
 linkcolor=blue, anchorcolor=blue,
 citecolor=blue, filecolor=blue,
 menucolor=blue, urlcolor=blue}

\def\beq{\begin{eqnarray}}
\def\eeq{\end{eqnarray}}

\usepackage[dvipsnames]{xcolor}

\def \beq {\begin{equation}}
\def \eeq {\end{equation}}
\begin{document}

\title{Observation of momentum dependent charge density wave gap in ${\mathbf{Eu}}{\mathbf{Te}_{4}}$}

\author{Iftakhar~Bin~Elius}\affiliation {Department of Physics, University of Central Florida, Orlando, Florida 32816, USA}
\author{Nathan~Valadez}\affiliation {Department of Physics, University of Central Florida, Orlando, Florida 32816, USA}
\author{Gyanendra~Dhakal}\affiliation {Department of Physics, University of Central Florida, Orlando, Florida 32816, USA}
\author{Volodymyr~Buturlim}\affiliation {Glenn T. Seaborg Institute, Idaho National Laboratory, Idaho Falls, Idaho 83415, USA}
\author{Sabin~Regmi}\affiliation {Department of Physics, University of Central Florida, Orlando, Florida 32816, USA}\affiliation {Center for Quantum Actinide Science and Technology, Idaho National Laboratory, Idaho Falls, Idaho 83415, USA}
\author{Dante~James}\affiliation {Department of Physics, University of Central Florida, Orlando, Florida 32816, USA}
\author{Peter~Radanovich}\affiliation {Department of Physics, University of Central Florida, Orlando, Florida 32816, USA}
\author{Matthew~Yankowitz}\affiliation{Department of Physics, University of Washington, Seattle, Washington, 98195, USA}\affiliation{Department of Materials Science and Engineering, University of Washington, Seattle, Washington, 98195, USA}
\author{Tetiana~Romanova}
\affiliation{Institute of Low Temperature and Structure Research, Polish Academy of Sciences, Ok\'{o}lna 2, 50-422 Wroc\l{}aw, Poland}
\author{Andrzej~Ptok}\affiliation{
Institute of Nuclear Physics, Polish Academy of Sciences, W. E. Radzikowskiego 152, PL-31342 Krak\'{o}w, Poland}
\author{Krzysztof~Gofryk}\affiliation {Center for Quantum Actinide Science and Technology, Idaho National Laboratory, Idaho Falls, Idaho 83415, USA}
\author{Dariusz~Kaczorowski}
\affiliation{Institute of Low Temperature and Structure Research, Polish Academy of Sciences, Ok\'{o}lna 2, 50-422 Wroc\l{}aw, Poland}
\author{Madhab~Neupane}
\email[e-mail: ]{madhab.neupane@ucf.edu}\affiliation{Department of Physics, University of Central Florida, Orlando, Florida 32816, USA}

\begin{abstract}
The occurrence of charge density wave (CDW) phenomena, particularly in low-dimensional rare-earth chalcogenides, has attracted substantial research interest.
Among these materials, EuTe$_4$, which features multiple Te layers and a single Eu-Te layer, serves as a promising platform to study the interplay between CDW order and 4$f$ electron configurations, including magnetism.
In this study, First-principles based density functional theory (DFT) calculations were carried out to investigate the electronic band structure modifications arising from CDW modulation.
Angle-resolved photoemission spectroscopy (ARPES) revealed the emergence of a CDW gap at the Fermi level, as well as hybridization-induced gap features at lower binding energies.
The low-lying CDW gap reaches its maximum along the $\overline{\Gamma}$--$\overline{\text{Y}}$ high-symmetry direction and a minimum along $\overline{\Gamma}$--$\overline{\text{X}}$, reflecting the anisotropic nature of the electronic structure.
We also performed low-temperature heat capacity measurements in applied magnetic fields near the N\'eel temperature (T$_N$~$\approx$ 6.9~K) to construct the magnetic phase diagram of EuTe$_4$.
This study provides valuable insight into the directional dependent evolution of the Fermi surface nesting induced CDW ordering, along with other observed gap openings within this system.
\end{abstract}

\maketitle

\section{Introduction}

Charge density waves (CDWs) are collective electronic states that have attracted significant interest, particularly in low-dimensional systems, due to their intricate interplay with superconductivity and magnetism~\cite{CDWreview, voit2000electronic, yokoya2001fermi, PRLCDW, Iyeiri_RTe3_2003, Zeng_CuIrCrTe_2022}.
Typically, a CDW is characterized by a periodic modulation of electronic charge density coupled with a lattice distortion, leading to a lowered translational symmetry. The emergence of a CDW usually involves the opening of an energy gap at specific regions of the Fermi surface (FS), driven primarily by FS nesting, where certain FS segments are connected by a common nesting vector.
This gap opening can significantly alter the electronic structure, potentially driving a metal-to-insulator transition. However, alternative mechanisms, such as structural instabilities, magnetic ordering, and electron-electron interactions, can also induce CDWs with characteristic wave vectors independent of FS nesting~\cite{CDWreview}.

In two-dimensional (2D) systems, CDW formation is typically explained using Peierls instability theory, which predicts FS nesting-driven gap openings and partial gapping of electronic states~\cite{peierls1996quantum, Pouget_pierls_instability_2015}.
In higher-dimensional systems (3D), FS nesting is generally imperfect, resulting in partial gap openings and thus metallic characteristics persist even in the CDW state~\cite{Gweon_SmTe3_PRL98, Garcia_LaTe2_2007, GdTe3_sabin}. A key distinction between one-dimensional (1D) and higher-dimensional CDW systems is the degree of FS gapping. Ideal nesting conditions in 1D systems typically result in a fully gapped FS and insulating behavior~\cite{CDWreview}, whereas 2D or 3D systems usually develop gaps only at specific nested regions, retaining metallic characteristics~\cite{Colona, sipos2008mott, Kidd, PRMEuTe4}.
Among the most extensively studied CDW materials are transition metal dichalcogenides (TMDCs, $TX_{2}$, $T =$ transition metals, $X =$ S, Se), which serve as prototypical platforms for exploring fundamental aspects of CDW physics~\cite{Rossnagel_2011, Ru_thesis, Xi2015, Soumyanarayanan2013}.
TMDCs exhibit a remarkable diversity in CDW phenomena due to their layered structures, tunable electronic dimensionality, and sensitivity to external parameters such as pressure, doping, and temperature.
These materials have provided critical insights into CDW phase transitions, gap opening mechanisms, and their interplay with superconductivity, making them ideal systems for investigating quantum phase transitions and correlated electronic states~\cite{Rossnagel_2011, Xi2015, Soumyanarayanan2013}.
More recently, quasi-2D rare-earth polytellurides, $R$Te$_{n}$ ($R =$ rare-earth, $1\!<\!n\!\leq \! 4$), have also attracted considerable attention due to their unique FS geometries, which are particularly conducive to nesting-driven CDW instabilities~\cite{PRMEuTe4, magnetic_CDW, Ru_thesis}.
The layered crystal structure of rare-earth polytellurides is characterized by consecutive Te square nets separated by corrugated $R$-Te slabs~\cite{PRMEuTe4}.
The origin of the CDW instability is primarily attributed to the Te $p$ orbitals derived from the planar Te sheets or the $R$-Te layers~\cite{PRMEuTe4, Garcia_LaTe2_2007, PrTe3}.
Although significant research has been conducted on rare-earth dichalcogenides and trichalcogenides, most of these materials exhibit only partial FS nesting, resulting in imperfect CDW-induced band folding~\cite{Shin, Kang, Garcia_LaTe2_2007, Brouet1, Brouet2, Kang1, Kogar2019, PrTe3}.
Angle-resolved photoemission spectroscopy (ARPES) studies on LaTe$_3$ have revealed CDW gaps ranging from $50$~meV to $200$~meV below the Fermi level.
Periodic lattice distortions associated with multiband CDW states have also been observed using scanning tunneling microscopy~\cite{NbSe2, NbTe2_arpes, NbSe2arpes}.
In the layered compound TbTe$_3$ multiprobe studies have revealed the coexistence of magnetic order and CDW phenomena~\cite{Schmitt_TbTe3_2008, TbTe3_02}.
DyTe$_3$ has demonstrated strong momentum-dependent electron-phonon coupling resulting multiple CDW phases, as evidenced by ARPES studies~\cite{DyTe3_01, DyTe3_02, DyTe3_03}.
Furthermore, GdTe$_3$ has been reported as a non-Fermi-liquid metal, exhibiting exceptionally high carrier mobility and linearly increasing quasiparticle scattering rate with increasing binding energy~\cite{GdTe3}.
Recent studies have shown that the CDW gap in GdTe$_3$ exhibits pronounced momentum dependence, with the maximum gap magnitude systematically varying as a function of the rare-earth element~\cite{GdTe3_sabin, Brouet2}.
Recently, EuTe$_4$ has attracted significant attention owing to its incorporation of both mono- and bilayer tellurium pseudo-square sheets, which give rise to multiple competing CDW phases and lead to a fully reconstructed FS~\cite{Wang, EuTe4_gedik, EuTe4_zhang, Rathore_EuTe4_kohn_anomaly_2023}.
Although EuTe$_4$ shares a common electronic band structure motif with other rare-earth polytellurides ($R$Te$_n$), a distinctive and uncommon characteristic is its complete transition from metallic to insulating behavior~\cite{Lee_PrTe2_CeTe2_2015, Gweon_SmTe3_PRL98, Garcia_LaTe2_2007, Schmitt_TbTe3_2008, PrTe3, Sarkar_late3_2023}.
The presence of localized Eu 4$f$ electronic states further enhances the complexity and interest in this material~\cite{Lv_EuTe4_2024}. While previous studies reported presence of a stripe like CDW at higher temperatures, a recent study reported the emergence of a spindle like, off axis CDW phase at low temperatures (below $4$~K)~\cite{xiao.2024}.
Despite extensive experimental and theoretical investigations, a comprehensive understanding of the electronic band structure of EuTe$_4$, particularly its connection to the underlying symmetries of the two- and three-dimensional Brillouin zones (BZs), remains incomplete.
In addition, the momentum-dependent evolution of the CDW gap has yet to be thoroughly explored.
Notably, beyond the low-energy CDW gap near the Fermi level, EuTe$_4$ also exhibits a second CDW gap at higher binding energies~\cite{Fan_2018, EuTe4_gedik, EuTe4_zhang}.

\indent In this work, we present a study of the low-temperature thermodynamic properties of single-crystalline EuTe$_4$. The compound was found to exhibit a clear paramagnetic (PM) to antiferromagnetic (AFM) transition at approximately 6.9~K in zero applied field, with the transition temperature ($T_\text{N}$) gradually suppressed upon increasing the field.
To gain insight into the underlying electronic structure, we carried out first-principles calculations.
Furthermore, ARPES measurements were performed to probe the momentum dependent electronic structure, with particular focus on the reconstructed FS, CDW gaps and their evolution across the BZ and temperature.
These ARPES results enabled a detailed investigation of the directional dependence of the CDW gap.

 \begin{figure*}[ht]
	\centering
	\includegraphics[width=\linewidth]{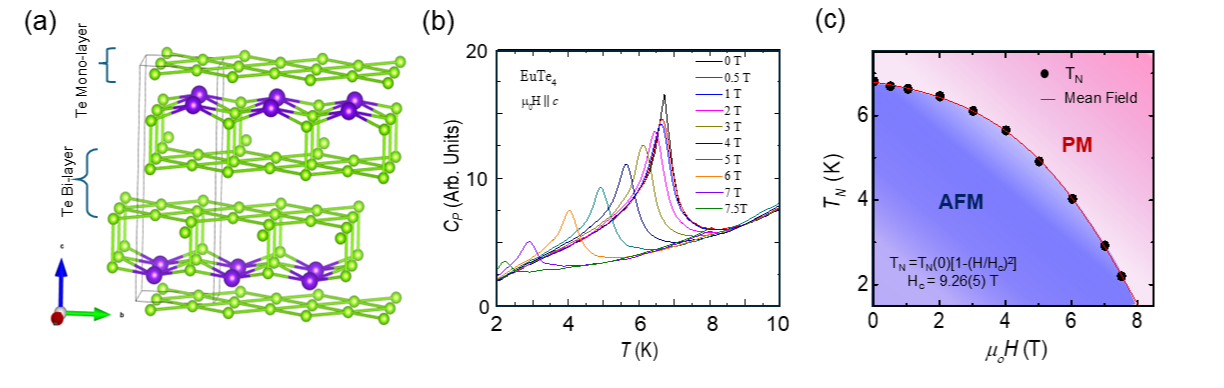}
	\caption{Crystal structure and field induced thermodynamic transport properties of EuTe$_{4}$. (a) Crystal structure of EuTe$_{4}$, the purple spheres represent Eu atoms, whereas green spheres denote Te atoms.(b) Temperature dependence of the specific heat measured in magnetic fields ranging from $0$~T to $7.5$~T (applied parallel to the c-axis) near the antiferromagnetic (AFM) to paramagnetic transition. (c)  Magnetic ($T-H$) phase diagram of EuTe$_4$ constructed from the N\'eel temperatures, fitted with mean field theory. The red dashed line represents a mean-field theoretical approximation (see text for details).
    }
\label{fig01}
\end{figure*}

\section{Methods}
\subsection*{Experimental techniques}
\subsubsection*{Crystal synthesis and characterization}
High quality single crystal pieces of EuTe$_{4}$ were grown via self-flux method, from a stoichiometric mixture of Eu and Te.
The mixture was placed in an alumina crucible and vacuum sealed in a quartz tube. Then heated to a temperature of $900^{\circ}$C, followed by slow cooling to $400^{\circ}$C over a period of 4 days. The flux was separated using a centrifuge.
Structural symmetry and phase purity of the samples were investigated using powder X-ray diffraction (XRD) analysis performed on finely ground crystals using a PANanalytical X’pert Pro diffractometer with Cu-K$_\alpha$ radiation.
The chemical composition was assessed using energy-dispersive X-ray (EDX) analysis conducted on a FEI scanning electron microscope with an EDAX Genesis XM4 spectrometer.
Furthermore, the single crystal pieces chosen for physical properties testing were inspected using an Oxford Diffraction X’calibur four-circle single-crystal X-ray diffractometer equipped with a CCD Atlas detector. Details of the structural characterization, CDW formation process, core-level spectrum etc. are presented in the SM article~A1 and SM Figs.~S1 and S2. 
The crystal structure presented in the main manuscript and the SM were visualized using software package VESTA~\cite{momma2011vesta}.

\subsubsection*{Bulk thermodynamic measurements}
Heat capacity measurements were carried out in the temperature interval of $2$--$300$~K in magnetic fields up to $7.5$~T using a Quantum Design DynaCool--14 Physical Property Measurement System (PPMS). A thin bar-shaped piece of single crystal with cross section $22.8 \times 134.5$~$\mu$m$^2$ was used for the measurements. The data were analyzed using OriginPro software package~\cite{originpro}.

\subsubsection*{Angle resolved photoemission spectroscopy}
ARPES measurements were performed at the Stanford synchrotron radiation light source (SSRL) beamline 5--2 equipped with a high efficiency DA30L electron analyzer. The energy resolution was better than $20$~meV and the angular resolution was better than $0.2^\circ$. The samples were cleaved $in~situ$ along the [001] Bragg plane and measured between $20$--$300$~K in a vacuum better than $10^{-10}$~Torr. The ARPES data analyses were performed using Igor Pro package~\cite{igorpro}. \\
\subsection*{Theoretical details}
First-principles calculations were carried out within the framework of density‑functional theory (DFT) using the projector‑augmented‑wave (PAW) formalism~\cite{blochl.94} as implemented in Vienna Ab initio Simulation Package (VASP)~\cite{kresse.hafner.94,kresse.furthmuller.96,kresse.joubert.99}.  
The exchange-correlation energy was described by the generalized‑gradient approximation (GGA) in the Perdew--Burke--Ernzerhof (PBE) parametrization~\cite{perdew.burke.96}.  
In the magnetic calculation the Eu 4$f$ electrons were treated as a valence states.
The correlation effects on the localized $f$ states we introduce by the DFT+U approach, proposed by Dudarev \textit{et al.}~\cite{dudarev.botton.98} (with $U = 5$~eV).
The energy cutoff for the plane-wave expansion was set to $350$~eV. 
Brillouin‑zone was probed using $16 \times 16 \times 4$ ${\bm k}$‑mesh within the Monkhorst--Pack scheme~\cite{monkhorst.pack.76}.  
As the convergence criterion of the optimization loop, we took the energy change below $10^{-6}$~eV and $10^{-8}$~eV for ionic and electronic degrees of freedom, respectively. 

\section{Results and Discussion}

\subsection*{Crystal structure and thermodynamic measurements}
EuTe$_{4}$ crystallizes in an orthorhombic crystal structure with symmetry $Pmmn$ (space group No.~59) at high temperature.
Which is a centrosymmetric nonsymmorphic space group.
The lattice constants were observed to be $a = 4.5119(2)$~\AA, $b = 4.6347(2)$~\AA, $c = 15.6747(10)$~\AA, $\alpha = \beta = \gamma = 90^{\circ}$. These values are in good agreement with previously reported values~\cite{PRMEuTe4, EuTe4_zhang}.
Upon cooling below the CDW transition temperature, the Te-layers undergo structural distortions to the $Pna2_1$ (space group No.~33)~\cite{PRMEuTe4}.  
In rare-earth ditellurides, the \mbox{$R$-Te} zig-zag chains are structurally enclosed between Te pseudo-square nets, which are rotated by $45^{\circ}$ with respect to the square nets directly coordinated to the rare-earth ($R$) atoms~\cite{Garcia_LaTe2_2007, Ru_thesis}.
In rare-earth tritellurides, a comparable structural motif is observed, where the $R$-Te layers are embedded between Te bilayers composed of two closely spaced, nearly square-planar Te sheets~\cite{ceTe3_K_Deguchi_2009, Sarkar_late3_2023, EuTe4_zhang}.
EuTe$_4$ distinguishes itself by hosting both the rotated Te monolayers and Te bilayers within each unit cell. In the $R$Te$_n$ family, such Te-derived layers are known to drive CDW instabilities via Fermi surface reconstructions~\cite{ceTe3_K_Deguchi_2009, Ru_thesis, Lee_PrTe2_CeTe2_2015, DyTe3_01}.
The simultaneous presence of both types of Te layers in EuTe$_4$ gives rise to a unique CDW phenomenon, emerging from the interplay and competition between the distinct nesting conditions and electronic instabilities associated with the monolayer and bilayer Te configurations~\cite{PRMEuTe4, EuTe4_gedik, EuTe4_zhang}.

Fig.~\ref{fig01}(a) presents the crystal structure of EuTe$_{4}$, depicting the layered structure consisting of the Eu-Te chain, Te-mono and Te-bilayers.
The non-CDW crystal structure of EuTe$_4$ is shown in Fig.~S1(a) and~S1(b) in the Supplementary Materials (SM)~\cite{SM}. 
Details of the structural transition and the resulting CDW (dimerization in the Te layers and oligomerization between the different Te layers) phase are provided in SM article A1~\cite{SM} and illustrated in SM Figs.~S1(c) and S1(d)~\cite{SM}.
The BZ and the core level spectrum of EuTe$_4$ are presented in Figs.~S2(a) and S2(b) in the SM~\cite{SM}, respectively.
The specific heat data for both zero field and fields applied parallel to the crystallographic $c$-axis are shown in Fig.~\ref{fig01}(b).
At zero field, a pronounced $\lambda$-shaped anomaly is observed near 6.9~K, indicative of a second-order phase transition from a PM to an AFM state.
Upon increasing the magnetic field, the transition peak systematically shifts to lower temperatures, and the height of the anomaly diminishes, reflecting the field-induced suppression of the AFM ordering.
The extracted $T_\text{N}$'s as a function of applied field are plotted in Fig.~\ref{fig01}(c).
The resulting phase boundary is well described by a mean-field model, $T_N(H) = T_N(0)[1 - \left(H/H_c \right)^2]$ (where $T_N(0)$ is the N\'eel temperature at zero applied magnetic field, $H$ is the applied magnetic field along c- direction, $H_c$ is the critical magnetic field at which the N\'eel temperature drops to zero), yielding a critical field of $\mu_0H_c = 9.26(5)$~T, beyond which the AFM order is completely suppressed.
The fit indicates a parabolic suppression of the transition temperature, suggesting a competition between magnetic field and exchange interactions. 
\subsection*{Theoretically calculated bulk band structure and experimental observation of CDW gap}
\begin{figure*} [ht]
	\centering
	\includegraphics[width=\linewidth]{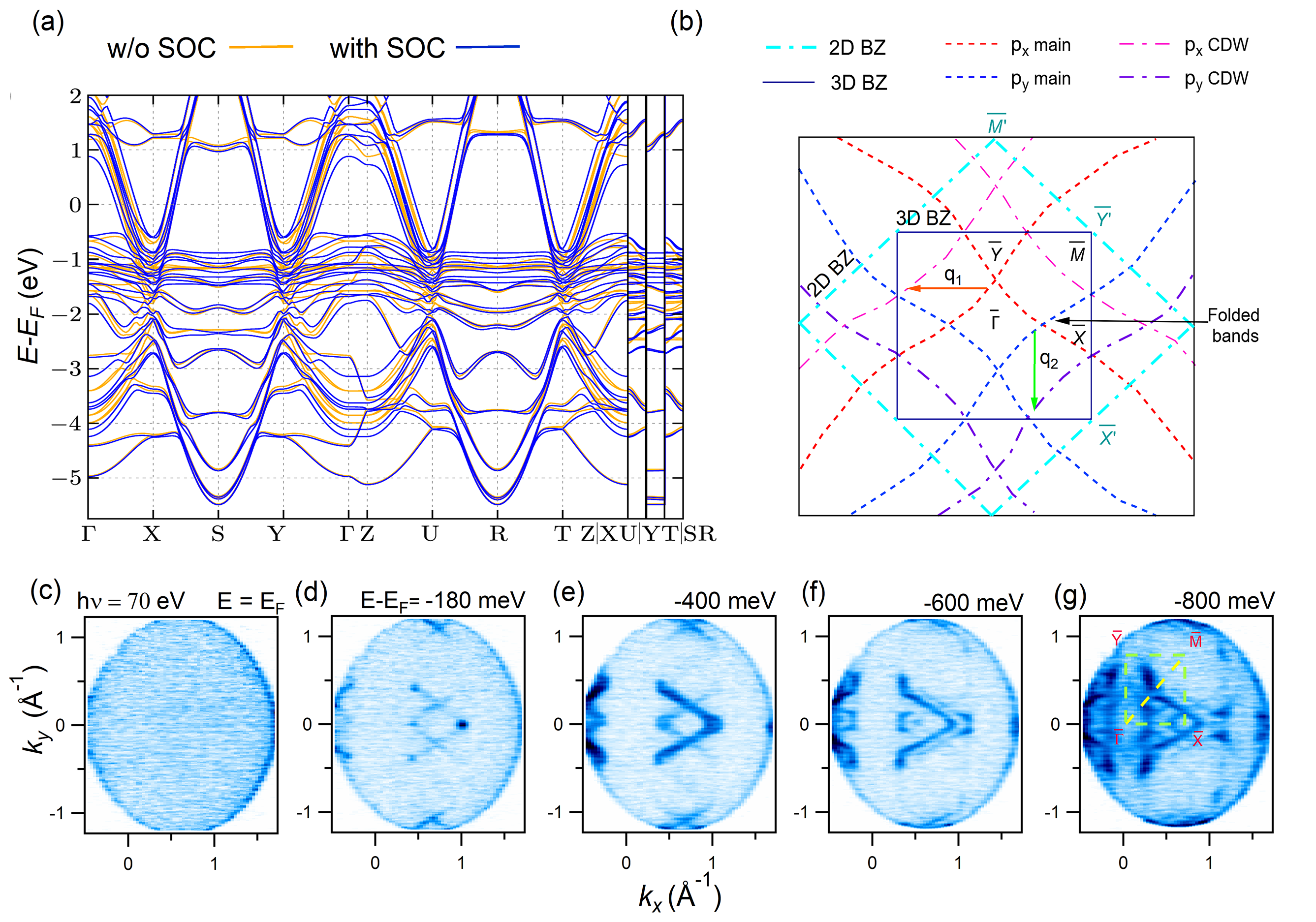}
	\caption{Electronic band structure of EuTe$_4$.
(a) Calculated band structure of EuTe$_{4}$ in the absence and presence of the spin--orbit coupling (yellow and blue lines, respectively), the Eu $f$ states were treated as valence states. FS map using a photon source of $70$~eV. 
(b) Schematic representation of constant energy contour (CEC) at 200~meV below the Fermi energy, exhibiting 2D and 3D Brillouin zones (BZs) and the nesting vectors ($q_1$ and $q_2$). 
(c) Experimentally observed Fermi surface (FS) and (d--g) CECs at different binding energies (indicated on the top of each plot). 
Measurements were performed at the SSRL beamline 5--2 at a temperature of $25$~K. 
\textbf{g} CEC at $E-E_{F}= -800~meV$ with high symmetry points $\overline{\Gamma}$, $\overline{\text{X}}$, $\overline{\text{Y}}$ and $\overline{\text{M}}$ marked, a quadrant of the BZ is indicated with the broken lines connecting the high symmetry points.}
\label{fig02}
\end{figure*}

In Fig.~\ref{fig02}(a), we present the bulk electronic band structure of EuTe$_4$ obtained from first-principles calculations, with and without the inclusion of spin--orbit coupling (SOC), shown in blue and yellow, respectively.
In these calculations, the Eu $4f$ electrons are treated as valence states. For comparison, an alternative calculation in which the $f$ electrons are treated as core states is provided in Fig.~S3 in the SM~\cite{SM}.
Additionally, bulk band structures calculated for the primitive, doubled, and tripled unit cells are shown in Fig.~S4 in the SM~\cite{SM}.
\begin{figure*}[ht] 
	\centering
	\includegraphics[width= \linewidth]{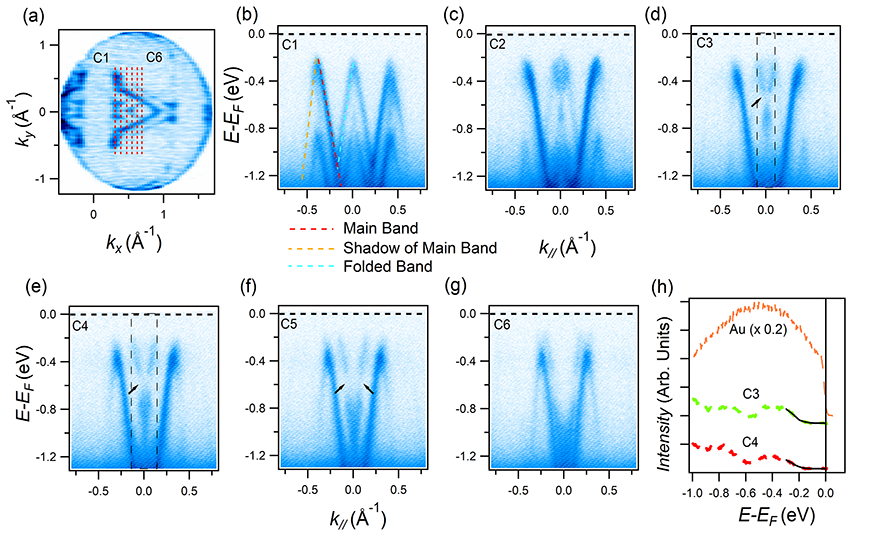}
	\caption{Momentum dependence parallel to $\overline{\Gamma}$-$\overline{\text{Y}}$ direction. (a) CEC map at $-800$~meV, positions of the dispersion maps parallel to $\overline{\Gamma}$-$\overline{\text{Y}}$ direction, are indicated with red dotted lines in the CEC map, indicated with C1 to C6, are presented in (b--g). 
    (h) Normalized intensities of cuts C3 and C4 (integrated over the range indicated by the boxes in respective figures) to discern the CDW and band modulation generated gaps (indicated by black arrows) presented with integrated spectra of a metallic (Au) sample to position the Fermi energy level. 
    ARPES measurements performed at SSRL beamline 5--2, at $T = 25$~K.}
\label{fig03}
\end{figure*}
\begin{figure*}[ht] 
	\centering
	\includegraphics[width= \linewidth]{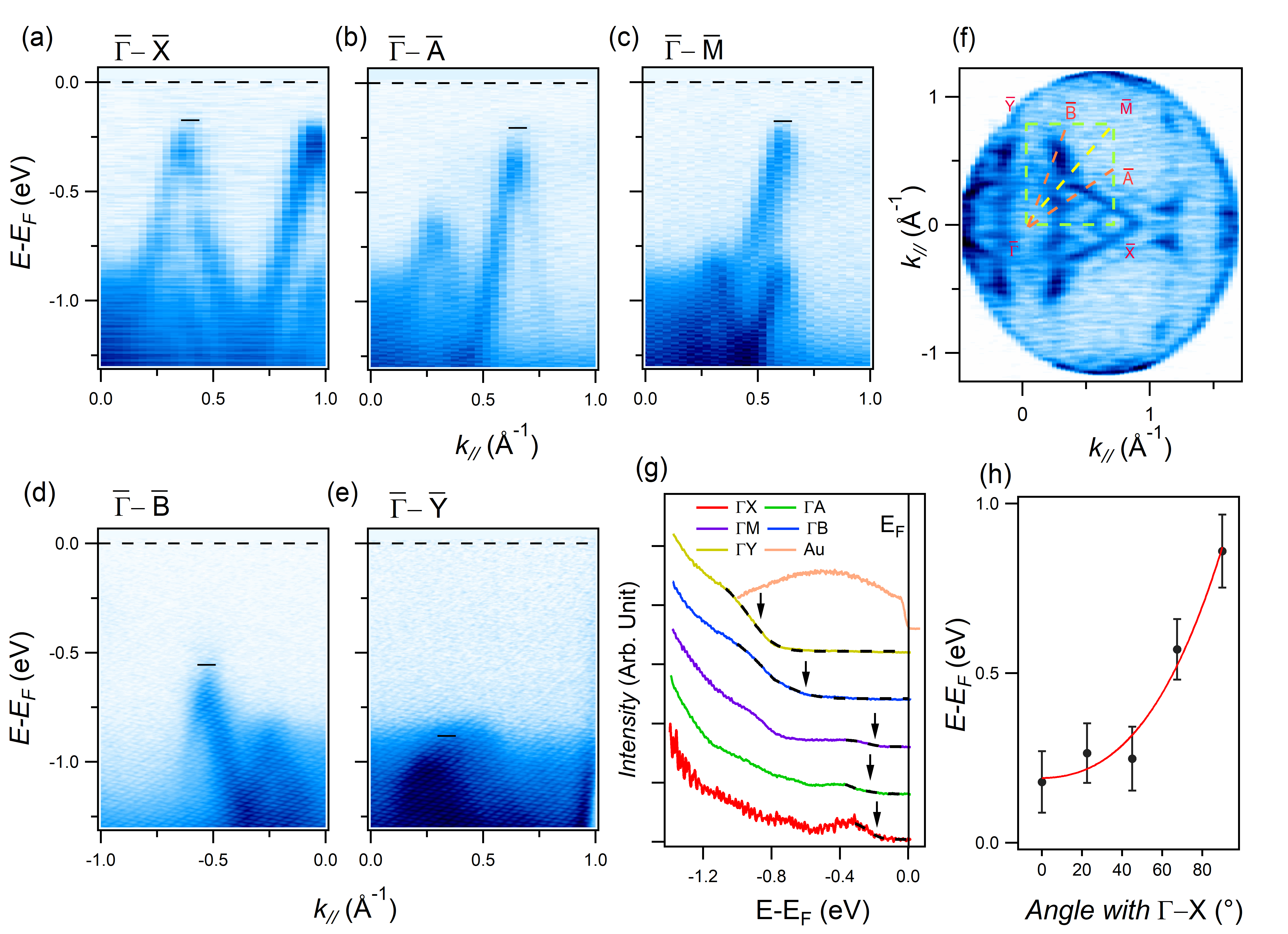}
	\caption{Radial evolution of the CDW gap in momentum space. (a--e) Dispersion maps along different radial positions from the $\overline{\Gamma}$-$\overline{\text{X}}$ direction. (f) CEC at $-800$~meV showing the positions of the radial directions centering the $\overline{\Gamma}$ point. (g) Energy distribution curves integrated in a window of $\pm 0.05$~eV at the leading edge positions along various directions. (h) CDW gaps with angular directions from the $\overline{\Gamma}$-$\overline{\text{X}}$. ARPES measurements were carried out at beamline 5--2 of the SSRL at a temperature of 25~K.}
\label{fig04}
\end{figure*}
A schematic constant energy contour (CEC) at a binding energy of 200~meV is illustrated in Fig.~\ref{fig02}(b), highlighting the primary (main) bands with $p_x$ and $p_y$ orbital characters (red and blue, respectively), along with the corresponding shadow bands (denoted in pink and purple).
The outer pseudo-square boundary (delineated by a broken turquoise line) indicates the two-dimensional BZ, which predominantly arises from the Te $5p_x$ and $5p_y$ orbitals (lying in the $ab$ plane).
In EuTe$_4$, the FS is composed of contributions from multiple Te layers and  presence of interlayer coupling via the Eu-Te layers~\cite{PRMEuTe4} (see the SM article A1 for a detailed discussion~\cite{SM}). 
However, the system is not purely 2D.
When accounting for the three-dimensional nature of the crystal structure, this pseudo-square 2D BZ is effectively reduced by a factor of $\sqrt{2}$ and rotated by $45^{\circ}$, as indicated by the solid navy-blue line in Fig.~\ref{fig02}(b).
Apart from this mixed dimensionality, another important aspect of the BZ is the anisotropic nature stemming from the reconstructed unit cell after CDW distortion. 
The measured FS map and CECs at various binding energies, acquired using a photon energy of 70~eV at a sample temperature of 25~K, are presented in Figs.~\ref{fig02}(c) and \ref{fig02}(d--g), respectively.
The complete absence of any spectral weight at the Fermi level across the FS suggests the presence of a persistent CDW gap at $E_{F}$.
As the binding energy increases, faint band features start to emerge at $E - E_{F} = -0.18$~eV, becoming more pronounced and clearly resolved within the $-400$ to $-800$~meV range.
At a binding energy of $-180$~meV, distinct features of both the primary and folded bands are observed along the $\overline{\Gamma}$---$\overline{\text{X}}$ and $\overline{\Gamma}$--$\overline{\text{M}}$ directions.
In contrast, along the $\overline{\Gamma}$--$\overline{\text{Y}}$ direction, no discernible photoemission signal is detected up to $-800$~meV, reflecting the inherent anisotropy arising from the orthorhombic symmetry ($a \neq b$) of the crystal, which reduces the FS symmetry from fourfold to twofold, originating from the anisotropic elongation of the unit cell (1$\times$3 after CDW distortion).
First-principles calculations reveal two prominent FS nesting vectors, $\mathbf{q_1} \! = \! \mathbf{b_1^*}/3$ and $\mathbf{q_2} \! = \! \mathbf{b_2^*}/3$ [see Fig.~\ref{fig02}b], consistent with the previous reports on this material~\cite{PRMEuTe4}.
At higher binding energies, the emergence of shadow bands associated with both, the main and the folded bands become increasingly evident.
Together, the main bands, folded bands, and their respective shadow bands generate a characteristic DNA-like pattern along the $\overline{\text{X}}$--$\overline{\text{M}}$ direction.
The magnitude of the CDW gaps along various high-symmetry directions is quantitatively evaluated in the subsequent subsection
For reference, Fig.~\ref{fig02}(g) displays a quadrant of the first BZ, marking the relevant high-symmetry points and directions used in the analysis.

Next, we examine the momentum dependence of the CDW gap along the $\overline{\Gamma}$--$\overline{\text{X}}$ high-symmetry direction, as shown in Fig.~\ref{fig03}.
Figures~\ref{fig03}(b--g) display a series of dispersion cuts (C1--C6) parallel to the $\overline{\Gamma}$--$\overline{\text{Y}}$ direction, marked by red dashed lines in Fig.~\ref{fig03}(a).
Cut C1 [Fig.~\ref{fig03}(b)] reveals the presence of the main bands, their shadow counterparts, the folded bands, and the corresponding shadow bands.
A consistent CDW gap of approximately 0.18~eV is observed near the Fermi level ($E_F$).
In dispersion cuts C3 through C5, an additional gap is observed within the folded bands at higher binding energies (shown by arrows in Figs.~\ref{fig03}(d--f)).
This higher-energy gap remains stationary across these momentum cuts, suggesting it originates from a distinct mechanism.
Notably, in some members of the $R$Te$_n$ family, including EuSbTe$_3$, two separate CDW gaps have been reported, each arising from different physical origins~\cite{Brouet2,Fan_2018}.\\
\indent To further elucidate the nature of these gap features, we present normalized energy distribution curves (EDCs) corresponding to cuts C3 and C4, extracted over the momentum windows indicated with the dashed boxes in Figs.~\ref{fig03}(d) and (e).
These EDCs confirm the robustness of the low-energy CDW gap and were fitted using a Fermi-edge function (for details of the Fermi edge fitting see article~A3 in the SM~\cite{arfken2012mathematical, GdTe3_sabin, SM}).
A more detailed visualization, including the dispersion cuts along $\overline{\Gamma}$--$\overline{\text{M}}$ and $\overline{\Gamma}$--$\overline{\text{X}}$, the second derivative of the $\overline{\Gamma}$--$\overline{\text{X}}$ cut, and the corresponding fitted EDCs, is provided in the Fig.~S5 in the SM~\cite{SM}.
Additionally, the suppressed spectral weight near $-500$~meV appears approximately at a distance of $0.3$~\AA$^{-1}$ from the $\Gamma$ point along the $\overline{\Gamma}$--$\overline{\text{M}}$ high symmetry direction (see Fig.~S6 in the SM~\cite{SM}). 

\subsection*{Radial momentum dependence of the gaps}
Subsequently, we explore the momentum dependence of the gaps moving radially around the $\overline{\Gamma}$ point at every $22.5^{\circ}$, along the $\overline{\Gamma}$--$\overline{\text{X}},~\overline{\Gamma}$-$\overline{\text{A}},~\overline{\Gamma}$--$\overline{\text{M}}, ~\overline{\Gamma}$-$\overline{\text{B}}$ and~$\overline{\Gamma}$-$\overline{\text{Y}}$ directions, as shown in Fig. \ref{fig04}f.
The above mentioned ARPES cuts are presented in the Fig.~\ref{fig04}(a--e).
To better quantify the evolution of the gap, the EDCs integrated over $\pm 0.05$~\AA$^{-1}$ are presented in Fig. \ref{fig04}(g).
The leading edges of the energy profiles are fitted with the Fermi edge function (see article~A3 in the SM~\cite{SM}) and indicated with black broken lines.
The CDW gap is ubiquitously present in all the momentum directions, reaching its maximum along $\overline{\Gamma}$--$\overline{\text{Y}}$ direction $\sim 780$~meV. 
The leading edges from $\overline{\Gamma}$--$\overline{\text{X}}$ to $\overline{\Gamma}$--$\overline{\text{B}}$ are contributions from $p_x$ and $p_y$ bands of Te- square planar layers. 
The low-lying bands along the $\overline{\Gamma}$--$\overline{\text{X}}$ direction are mostly from Te- atoms of Eu-Te zigzag layer.
A comparative picture of the experimentally observed CDW gaps is presented in Fig.~\ref{fig04}(h). 
To better visualize the modulation driven gap and quantifying their extents, cuts along CT1 and CT2 are presented in Fig.~S6 in the SM~\cite{SM}.
The EDCs along these cuts at the momentum positions corresponding to these hybridized gap establishes the same extent, relative position with respect to $\overline{\Gamma}$ point, and similar width. 
The maximum intensity points from the dispersion maps (of different directions from the $\overline{\Gamma}$--$\overline{\text{X}}$ direction) were identified to calculate the Fermi velocities along the main, shadow and folded bands.
Unlike $R$Te$_3$ materials, no bilayer splitting was observed over the momentum space. 
ARPES spectrographs, measured at other incident photon energies or other temperatures (within the upper CDW critical temperature) exhibited similar CDW driven FS reconstruction and the modulation driven gap (see the radial cuts of ARPES spectra, their EDCs measured with $75$~eV photon energy at 35~K presented in Fig.~S7 in the SM~\cite{SM}).
Around at 1.4~eV below the Fermi energy, Eu 4$f$ originated nondispersive flat bands are observed (see Fig.~S7 in the SM~\cite{SM}). 
These flat bands are in good agreement with our theoretical calculation presented in Fig.~\ref{fig02}(a). 
 
The existence of CDW stemming from mono- and bi- Te-layers in EuTe$_4$ sets this material into a distinctive class of its own among the $R$Te$_n$ family. The CDW distortion in mono- and bi-layer Te nets, hosting the CDW phenomena is triggered in different temperatures, which is reflected in the hysteretic nature in the thermodynamic and electrical transport measurements~\cite{PRMEuTe4, EuTe4_gedik, EuTe4_zhang}. Moreover, existence of Eu 4$f$ electrons interacting with CDW phenomena, collectively made EuTe$_4$, a wonderful platform to study the interplay of long range correlation and CDW phase. In $R$Te$_n$ materials, Te 5$p$ bands hosted by the weakly interacting Te-layers construct the electronic structure in the vicinity of FS. In RTe$_2$ and RTe$_3$, consecutively Te mono and bi-layers form the electronic structure near the E$_F$. Below, the CDW transition temperature, the Te-nets go through periodic distortions, forming Te-trimers, particularly, in case of EuTe$_4$, the periodic distortion extends into a (1$\times$3$\times$2) superstructure. The pseudo-square nature of the $ab$-plane ($\sim$0.12~\r{A}) and the unequal superlattice cell parameters, results in the strong direction dependent nature of the band folding or CDW gap. Similar to other iso-structural $R$Te$_n$ materials, the band signature in EuTe$_4$ is two fold symmetric instead of a fourfold symmetry~\cite{GdTe3, GdTe3_sabin}. But most of the $R$Te$_2$ or $R$Te$_3$ materials, do not undergo complete metal to insulator transition as not all the bands undergo CDW reconstruction. In EuTe$_4$, a unique global CDW gap opening was observed in our ARPES dispersion maps, radially taken along $\overline{\Gamma}$--$\overline{\text{X}}$ to  $\overline{\Gamma}$--$\overline{\text{Y}}$ (see Fig.~\ref{fig04}), from lowest to highest in terms of gap size, respectively. As theoretical calculations are normally not capable of capturing the CDW phenomena, ARPES based experimental probes were instrumental in finding the direction dependence of the CDW gap in momentum space.  Despite the band features till discussed are attributed to the Te-layers, several studies have established that there are possible superexhange interactions between the $R$- atoms in $R$-Te layer, mediated by the Te-layers~\cite{ceTe3_K_Deguchi_2009}. This exchange coupling plays a key role in the formation of the CDW-induced band folding. The strong presence of the folded bands (in the CECs at higher binding energies and cuts C1$\sim$ C6, presented in Fig.~\ref{fig03}) hints this type of interaction in case of EuTe$_4$~\cite{GdTe3}.
A modulation driven gap opening was also observed at higher binding energies (visualized in the cuts C2$\sim$C5 and in the $\overline{\Gamma}$--$\overline{\text{X}}$  direction), in addition to the CDW driven band gap opening.
This band gap, observed in the folded bands, is omnipresent in ARPES measurements performed at other photon energies or temperatures. The low-lying CDW gap exhibits a progressive increase in magnitude with decreasing temperature, whereas the higher binding energy gap diminishes with increasing temperature and becomes fully suppressed at and above 300~K.
To provide a quantitative comparison, the evolution of both gap sizes as a function of temperature is plotted in Fig.~S8 in the SM~\cite{SM}.
Given that the primary CDW transition temperature in EuTe$_4$ is significantly higher than 300~K, the persistence of the low-energy CDW gap is consistent with expectations.
In contrast, the temperature-sensitive closure of the higher binding energy gap (see SM Fig.~S8(a)~\cite{SM}) suggests a fundamentally different origin. At elevated temperatures, thermal fluctuations increasingly inhibit lattice scattering processes (which gives rise to the folded bands).  This is believed to be responsible for the disappearance of the modulation-induced gap at higher binding energies (which is in alignment with the previous reports on EuTe$_4$~\cite{zhang.wu.22,EuTe4_zhang} and similar systems~\cite{Brouet2, Fan_2018}).

\section{Conclusion}
In summary, we performed heat capacity measurements in zero field and under applied magnetic fields up to 7.5~T.
A pronounced peak is observed at 6.9~K, corresponding to the AFM phase transition in this material.
The magnetic order is gradually suppressed by external magnetic fields and vanishes at around 9~T, indicating a field-driven suppression of antiferromagnetism.
Based on these results, we constructed a magnetic phase diagram of EuTe$_4$.
High-resolution ARPES measurements were carried out on EuTe$_4$ to resolve the detailed momentum dependence of the CDW and hybridization gaps, employing radial cuts centered at the $\overline{\Gamma}$ point and systematic scans along and parallel to the high-symmetry directions and relevant band dispersions.
Strong momentum dependent main, folded and shadow bands originate from $\sim$200~meV below the Fermi level along $\overline{\Gamma}$-$\overline{\text{X}}$ direction and reaches minimum along $\overline{\Gamma}$-$\overline{\text{Y}}$.
Overall, EuTe$_4$ offers an excellent platform to study the effect of 4$f$ electrons, their magnetism and CDW order.\\

\vspace{2ex}

\section*{acknowledgments} 
\vspace{-0.1 cm}
M.N. acknowledges the support from the National Science Foundation (NSF) CAREER Award No.~DMR-1847962, and the NSF Partnerships for Research and Education in Materials (PREM) Grant No.~DMR-2424976. 
Work at INL was supported by the US Department of Energy, Basic Energy Sciences, Materials Sciences, and Engineering Division.
T.R. and D.K. are supported by the National Science Centre (NCN, Poland) under research grant 2021/41/B/ST3/01141. 
A.P. acknowledges the support by National Science Centre (NCN, Poland) under project No. 2021/43/B/ST3/02166. 
This research utilized resources of the Stanford synchrotron radiation lightsource (SSRL) BL5--2. 
We express our gratitude to Dr.~Donghui Lu and Dr.~Makoto Hashimoto for providing valuable support with the beamline at the SSRL. 
We acknowledge Dr.~Anup Pradhan~Sakhya, Milo~Sprague and Mazharul~Islam~Mondal for their help during the experiment and fruitful discussion during the manuscript preparation.

\end{document}